\documentclass[a4paper]{article}   %
\pdfoutput=1

\newif\ifmytwocolumn
\mytwocolumnfalse
\relax

\usepackage{hyperref}
\usepackage[comma]{natbib}
\newcommand{\doi}[1]{\textsc{doi}: \href{http://dx.doi.org/#1}{\nolinkurl{#1}}}
\setcitestyle{square,numbers}

\usepackage{graphicx}
\usepackage{siunitx}
\usepackage{authblk}
\usepackage{xcolor} 
\usepackage{tikz}
\usetikzlibrary{shapes}
\usepackage[normalem]{ulem}  %
\usepackage{amsmath}

\usepackage{textcomp}
\usepackage{gensymb}
\usepackage{needspace}

\usepackage{stackengine}
\stackMath
\newcommand\tsup[2][2]{%
	\def\useanchorwidth{T}%
	\ifnum#1>1%
	\stackon[-1.3ex]{\tsup[\numexpr#1-1\relax]{#2}}{\mathchar"307E}%
	\else%
	\stackon[-1ex]{#2}{\mathchar"307E}%
	\fi%
}
\newcommand{\doubletilde}[1]{\tsup[2]{#1}}

\date{}

\usepackage{lineno}

\ifmytwocolumn
\usepackage[outer=2.59cm, inner=2.54cm]{geometry}
\else
\usepackage[outer=4.15cm, inner=4.15cm]{geometry}
\fi
\usepackage{marginnote}

\newcommand{\TODOobsolete}[1]{}
\newcommand{\TODOdone}[1]{}
\newcommand{\TODOdrop}[1]{}

\newcommand{\ecm}{\textit{e}\,cm}

\newcommand{\eg}{\textit{e.g.\ }}

\newcommand{\ie}{\textit{i.e.\ }}

\newcommand{\comagnetometer}{co-mag\-n\-eto\-meter}

\newcommand{\smallAffilText}[1]{{\small #1}}

\begin{document}
\bibpunct{\,\,[}{]}{,}{n}{}{;}    %

\newlength{\jkfigurewidth}	
\ifmytwocolumn
\setlength{\jkfigurewidth}{\columnwidth}
\newcommand{\myTikzStrechFactor}{0.54}	
\else
\setlength{\jkfigurewidth}{85mm}  %
\setlength{\jkfigurewidth}{95mm}  %
\newcommand{\myTikzStrechFactor}{0.6}	
\fi

\title{Data blinding for the nEDM experiment at PSI}
\author[1,2]				{\textcolor{black} {N. J. Ayres}}    %
\author[3] 				{\textcolor{black} {G. Ban}}
\author[4]   				{\textcolor{black} {G. Bison}}    %
\author[5] 				{\textcolor{black} {K. Bodek}}
\author[1,4,6]		{\textcolor{black} {V. Bondar}}
\author[7] 				{\textcolor{black} {E. Chanel}}
\author[4]   				{\textcolor{black} {P.-J. Chiu}}
\author[8]   			{\textcolor{black} {C. Crawford}}
\author[4]   				{\textcolor{black} {M. Daum}}
\author[1]  				{\textcolor{black} {S. Emmenegger}}
\author[9]				{\textcolor{black} {L. Ferraris-Bouchez}}  %
\author[3] 				{\textcolor{black} {P. Flaux}}
\author[10,$\dagger$]	{\textcolor{black} {Z. Gruji\'c}}
\author[2]  			{\textcolor{black} {P.G Harris}}   %
\author[4]   				{\textcolor{black} {N. Hild}}
\author[3] 				{\textcolor{black} {J. Hommet}}   %
\author[4,6,10]{\textcolor{black} {M. Kasprzak}}
\author[9]				{\textcolor{black} {Y. Kermaidic}}%
\author[1,4]			{\textcolor{black} {K. Kirch}}  %
\author[1,4] 			{\textcolor{black} {S. Komposch}}
\author[11] 				{\textcolor{black} {A. Kozela}}
\author[$\ast$,1]		{\textcolor{black} {J. Krempel}} %
\author[4]   				{\textcolor{black} {B. Lauss}}
\author[3]				{\textcolor{black} {T. Lefort}}   %
\author[3]				{\textcolor{black} {Y. Lemiere}}  %
\author[9]				{\textcolor{black} {A. Leredde}}    %
\author[1,4]   			{\textcolor{black} {P. Mohanmurthy}}
\author[4]   				{\textcolor{black} {A. Mtchedlishvili}}
\author[3] 				{\textcolor{black} {O. Naviliat-Cuncic}}
\author[4]   				{\textcolor{black} {D. Pais}}
\author[7] 				{\textcolor{black} {F. M. Piegsa}}
\author[9]				{\textcolor{black} {G. Pignol}}  %
\author[1] 				{\textcolor{black} {M. Rawlik}}
\author[9]				{\textcolor{black} {D. Rebreyend}}   %
\author[4]   				{\textcolor{black} {I. Rien\"acker}}
\author[12]				{\textcolor{black} {D. Ries}}     %
\author[13,14]      	{\textcolor{black} {S. Roccia}}
\author[5] 				{\textcolor{black} {D. Rozpedzik}}
\author[4]   				{\textcolor{black} {P. Schmidt-Wellenburg}}
\author[15]				{\textcolor{black} {A. Schnabel}}   %
\author[9]				{\textcolor{black} {R. Virot}}   %
\author[10]			{\textcolor{black} {A. Weis}}  %
\author[6,$\ddagger$]{\textcolor{black}{E. Wursten}}  %
\author[5]				{\textcolor{black}{J. Zejma}}    %
\author[4]   				{\textcolor{black} {G. Zsigmond}}

\affil[1]{\smallAffilText{Institute for Particle Physics and Astrophysics, ETH Z{\"u}rich, Z{\"u}rich, Switzerland}}
\affil[2]{\smallAffilText{Department of Physics and Astronomy, University of Sussex, Falmer, Brighton, UK}}
\affil[3]{\smallAffilText{LPC Caen, ENSICAEN, Université de Caen, CNRS/IN2P3, Caen, France}}
\affil[4]{\smallAffilText{Paul Scherrer Institute, Villigen, Switzerland}}
\affil[5]{\smallAffilText{M. Smoluchowski Institute of Physics, Jagiellonian University in Krakow, Poland}}
\affil[6]{\smallAffilText{Instituut voor Kern- en Stralingsfysica, Katholieke Universiteit Leuven, Leuven, Belgium}}
\affil[7]{\smallAffilText{University of Bern, 	Albert Einstein Center for Fundamental Physics, 	Laboratory for High Energy Physics, 	Bern, Switzerland}}
\affil[8]{\smallAffilText{Department of Physics and Astronomy, University of Kentucky, Lexington, USA}}
\affil[9]{\smallAffilText{Univ. Grenoble Alpes, CNRS, Grenoble INP, LPSC-IN2P3, Grenoble, France}}
\affil[10]{\smallAffilText{Physics Department, University of Fribourg, Fribourg, Switzerland}}
\affil[11]{\smallAffilText{H. Niewodniczanski Institute of Nuclear Physics, Polish Academy of Sciences, Krakow, Poland}}
\affil[12]{\smallAffilText{Department of Chemistry - TRIGA site, Johannes Gutenberg University Mainz, Mainz, Germany}} %
\affil[13]{\smallAffilText{CSNSM, Universit\'e Paris Sud, CNRS/IN2P3, Universit\'e Paris Saclay, Orsay-Campus, France}}
\affil[14]{\smallAffilText{Institut Laue-Langevin, Grenoble, France}}
\affil[15]{\smallAffilText{Physikalisch Technische Bundesanstalt, Berlin, Germany}\vspace{0.5\baselineskip}}

\affil[$\dagger$]{\smallAffilText{Present address: Institute of Physics Belgrade, Belgrade, Serbia}}
\affil[$\ddagger$]{\smallAffilText{Present address:  CERN, Geneva, Switzerland}\vspace{0.5\baselineskip}}

\affil[$\ast$]{Corresponding author: Jochen.Krempel@phys.ethz.ch}
\maketitle
\begin{abstract}
Psychological bias towards, or away from, a prior measurement or a
theory prediction is an intrinsic threat to any data analysis. While
various methods can be used to avoid the bias, \eg actively not
looking at the result, only data blinding is a traceable and thus
trustworthy method to circumvent the bias and to convince a public
audience that there is not even an accidental psychological bias.

Data blinding is nowadays a standard practice in particle physics, but 
it is particularly difficult for  experiments searching for the neutron 
electric dipole moment (nEDM), as several cross measurements, in particular of 
the magnetic field, create a self-consistent network into which it is hard to
inject a fake signal.

We present an algorithm that modifies the data without influencing the experiment.
Results of an automated analysis of the data are used to change the recorded spin state of a few neutrons of each measurement cycle.

The flexible algorithm is applied twice to the data, to provide different data to various analysis teams.
This gives us the option to sequentially apply various blinding offsets for separate analysis steps with independent teams.
The subtle modification of the data allows us to modify the algorithm and to produce a re-blinded data set without revealing the blinding secret.
The method was designed for the 2015/2016 measurement campaign of the nEDM experiment at the Paul Scherrer Institute. 
However, it can be re-used with minor modification for the follow-up experiment n2EDM, and may be suitable for comparable efforts.
  \end{abstract}

\section{Introduction}
The electric dipole moment (EDM) of the neutron is a fundamental observable in particle physics
 that may directly relate to the observed dominance of matter over antimatter in the Universe.  
It has been sought after for almost seven decades, with ever-improving sensitivity,
 but to date all experimental results have been compatible with zero\cite{nEDM-Review-Lamoreaux_2009,nEDM-limit-PhysRevLett.97.131801-2006,nEDM-limited-revised-2015-PhysRevD.92.092003,nEDM-result-PNPI-Serebrov-2015,Abel2020PRL}.
However, many beyond the Standard Model theories naturally predict values that are close to current experimental sensitivities%
  \cite{nedm-models-vs-experiments-ELLIS198933,Morrissey2012NJP,Engel2013PPNP}.  
Thus, depending upon their outlook, scientists analysing the data from EDM experiments may be biased unintentionally
 towards a result that favours their own expectations of either seeing, or not seeing, a statistically significant signal. 
Data blinding removes this psychological bias and,
 if applied properly, does not introduce any other bias.
In experimental particle physics, blinding has been used quite commonly for many years\cite{KleinRoodmanBlindAnalysisInNuclearAndParticlePhysics},
 but to date it has not been applied to any neutron EDM measurement.

In general, at least two different types of blinding can be distinguished:
 \begin{enumerate}
 \item{} Data corresponding to a region of interest is withheld from the analysis team,
           or, correspondingly, ``fake'' events can be added to obscure the signal.  
        This is often the case in discovery experiments.
        See, \eg (not the latest but representative) searches for rare decays\cite{MEG},
           dark matter\cite{Xenon100} or gravitational waves\cite{GravitationalWaves}.
 \item{} For precision experiments the observable of interest can be scaled by an unknown factor\cite{mulan}, or in some cases, an unknown offset can be added to the observable\cite{Vacheret2005}. 
\end{enumerate}
The latter is obviously applicable to EDM experiments,
 and it is the approach that we have adopted for the nEDM experiment at the Paul Scherrer Institute (PSI)\cite{Roccia-nEDM-data-taking-strategy-2018}.  
In deciding to modify the observable, one can choose to do so either by changing an aspect of the experiment itself, or by modifying the data {\em post hoc}.  
The latter has the advantage that it does not change or corrupt the experiment, and a hidden set of original data can be stored securely. 
Thus, if the blinding were to affect the data quality in any way, \eg by reducing the sensitivity,
 or by introducing a new bias, the original data can still be used
  knowing that the final result is unaffected by any systematic effects that may have been introduced through blinding.

\section{Experimental overview}
\begin{figure}[bt]
	\centering
	\includegraphics[width=\jkfigurewidth]{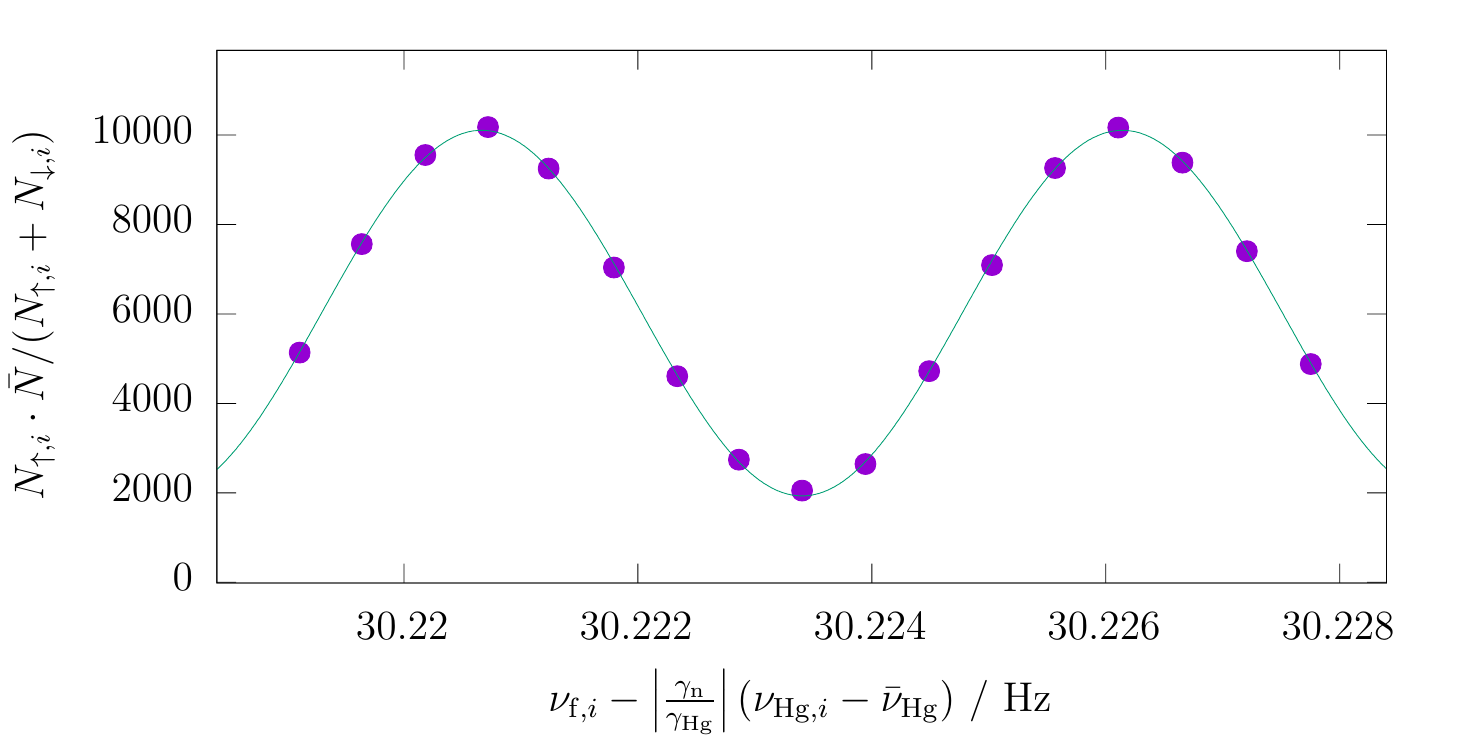}
	\caption{Measured neutron counts plotted versus spin-flip frequency.
		Both are corrected for fluctuations as indicated in the axis labels.
		A sine with offset is fitted to the data points.
		$\bar{\nu}_\mathrm{Hg}$ is the average reading of the mercury \comagnetometer.
	    Both averages used in this plot are calculated from the measurements shown in this graph.}
	\label{fig:ramsey-pattern-Nup}
\end{figure}

In nEDM experiments the target observable is
  the dependence of the neutrons' Larmor precession frequency upon an applied static electric field\cite{Roccia-nEDM-data-taking-strategy-2018}.
A set of ultracold neutrons (UCN) is polarized and then stored in a volume within a stable and highly uniform magnetic field.  
In most experiments to date the frequency measurement is based on Ramsey's technique of separated oscillatory fields. 
Two spin-flip pulses which induce each a $\pi/2$ flip through transverse oscillating magnetic fields are applied. 
The two pulses must have a well-known phase relation and were kept in phase at all times during our data taking.
Between these pulses the neutron spins can precess freely.
If the spin-flip frequency is perfectly in resonance with the Larmor frequency of the neutrons,
 they will have undergone a 
 $\pi$ spin-flip by the end of the procedure. 
If not, the accumulated phase difference, a highly sensitive measure of the difference between the Larmor and reference frequencies, 
   will result in a partial spin-flip.
The neutrons are then counted in a spin-sensitive detector.
By repeating such measurements while scanning the reference frequency, 
  and plotting the final neutron spin state versus that frequency,
  a Ramsey fringe pattern emerges as plotted in fig.\,\ref{fig:ramsey-pattern-Nup}. 
For a non-zero EDM value the pattern will shift when the electric field direction is reversed, which is done periodically.

Typically an entrance spin-flipper is present, in our case called SF1,
 which will invert the initial neutron spin orientation if employed.
Alternating its state is used to investigate the influence of systematic effects.
Regular changes of the magnetic field orientation and a variation of the magnetic field gradient serve the same purpose.

During the 2015/16 data taking campaign at PSI\cite{Roccia-nEDM-data-taking-strategy-2018} we detected up to \num{20000} UCN after having stored them in a volume of 22 litres
  within a highly uniform magnetic field of approximately \SI{1}{\micro T}.
Each of such single measurements, called ``cycle'', was composed of two $\pi/2$ spin-flip pulses of a frequency of $\nu_{\mathrm{F},i} \approx \SI{30}{Hz}$ applied for a duration of $t=\SI{2}{s}$ each
  and a free precession period of $T=\SI{180}{s}$
  before the UCN were counted.
  The detector simultaneously measured spin-up/down neutrons in two branches.
  Each consisted of a controllable spin-flipper, a magnetized spin-analysing foil,
    and \textsuperscript{6}Li based neutron detectors that were read out via photomultiplier tubes\cite{AfachHelaine-USSA,Ban-NANOSC-2016}.
    Timestamp, integrated charge and detector channel of every event were recorded in the data files.
A set of consecutive cycles carried out with a stable magnetic field configuration, but with variation of the applied electric field, was called a ``run''. 
During a run, lasting for up to several days, and typically consisting of several hundred
cycles, the configuration of the spin-flippers in the detector branches was reversed every four cycles,
 and the entrance spin-flipper status (the spin orientation of the neutrons let in) was changed every 112 cycles. 
Within those runs, eight cycles with no electric field were followed by 48 cycles with high voltage of a given polarity. 
Thus, both electric field polarities were applied between each change of the entrance spin-flipper state.

\subsection{Formal description}

The $\pi/2$ spin-flips of frequency $\nu_{\mathrm{F},i}$, that are applied in a particular cycle $i$, cause a change of polarization described by the phase 
\begin{equation}
\phi_i = \frac{(\nu_{\mathrm{F},i}-\nu_{\mathrm{L}})}{\Delta\nu}\pi \ ,
\label{eq:phi_i}
\end{equation}
where $\nu_{\mathrm{L}}$ is the Larmor frequency
 and the fringe width $\Delta\nu$ is  
\begin{equation} 
\Delta \nu = \frac{1}{2\left(T+4\,t/\pi \right)} \ ,
\label{eq:DeltaNu}
\end{equation} 
with $T$ and $t$ being respectively the free-precession time
and the duration of each spin-flip pulse.

The true numbers of neutrons $N^\prime_{\uparrow,i}$ and $N^\prime_{\downarrow,i}$ for the particular spin states
 that are in the precession chamber right before guiding them to the detector are
\begin{eqnarray}
N^\prime_{\uparrow,i} &=& \frac{N^\prime_i}{2} \left( 1 - \alpha^\prime \cos\phi_i \right)  \label{eq:ramsey_curve-true_neutrons-SpinUp}\\
N^\prime_{\downarrow,i} &=& \frac{N^\prime_i}{2} \left( 1 + \alpha^\prime \cos\phi_i \right) \ ,  
\label{eq:ramsey_curve-true_neutrons-SpinDown}
\end{eqnarray}
where $N^\prime_i$ is the total number of neutrons in the chamber after the precession and $\alpha^\prime$ describes the true visibility after the precession, but has a negative sign in case SF1 is enabled.

The neutrons will then fall through a polarisation analyser with spin se\-lec\-tiv\-i\-ties $p_\mathrm{A}$ and $p_\mathrm{B}$ to the detectors that operate with efficiencies $\epsilon_\mathrm{A}$ and $\epsilon_\mathrm{B}$.
Thus, the numbers of measured neutrons are
\begin{eqnarray}
N_{\uparrow,i} &=&  \big( N^\prime_{\uparrow,i} \, p_\mathrm{A} + N^\prime_{\downarrow,i} (1-p_\mathrm{A}) \big) \epsilon_\mathrm{A} \label{eq:detectedNeutrons-SpinUp}\\
N_{\downarrow,i} &=& \big( N^\prime_{\downarrow,i} \, p_\mathrm{B} + N^\prime_{\uparrow,i} (1-p_\mathrm{B}) \big) \epsilon_\mathrm{B} \ .
\label{eq:detectedNeutrons-SpinDown}
\end{eqnarray}
In this model the efficiency of the spin-flippers is neglected since it is very close to 1. 

``Spin-up''~($\uparrow$) refers to neutrons
with the spin polarisation anti-parallel to the magnetic field $B_0$,
and therefore with the magnetic moment parallel to the field.
They are also known as ``high field seekers''.
When SF1 is off, this is the state in which they enter the bottle 
 and is thus their state before the Ramsey sequence is applied.

A fit of eqs.\,\eqref{eq:phi_i}--\eqref{eq:detectedNeutrons-SpinDown} to the data yields the Larmor frequency. 
The steepest part of the slope, \ie where $\phi_i\approx \pm 90\degree$, is most sensitive to variations in frequency.
Thus the spin-flip frequencies were configured to operate sequentially at four working points (see points in fig.\,\ref{fig:shift-ramsey-pattern}) close to this condition
  with a detuning of about 5\% of the fringe width in order to have some sensitivity also to further experimental parameters such as the visibility
  and the asymmetry of the detector efficiency.
 
In the presence of an nEDM $d$ and an applied electric field $\vec{E}$ collinear to the magnetic field $\vec{B}$, the resonant frequency $\nu_{\mathrm{L}}$ shifts by 
\begin{equation}
\delta \nu = 2\,d\,  \vec{E} \cdot \frac{\vec{B}}{\left|\vec{B}\right|} /h \ ,
\label{eq:delta_nu-resonanceshift-nedm}
\end{equation}
with $h$ being the Planck constant. Note that the $\vec{B}/\left|\vec{B}\right|$ term is required only to obtain the proper sign.

Unfortunately, any change of the amplitude of the magnetic field causes a corresponding change of the Larmor frequency. 
We used a mercury \comagnetometer\ to correct for magnetic-field fluctuations
  by taking the ratio of the measured frequencies $\mathcal{R}=\nu_\mathrm{n}/\nu_\mathrm{Hg}$\cite{BAKER2014184}. 
Correspondingly, eq.\,\eqref{eq:phi_i} is altered as shown in eq.\,\eqref{Lamor-frequency-vs-mercury-and-gradient-phase}.  
However, although the (thermal) mercury atoms populate the storage cell rather uniformly,
  the UCN have such low kinetic energies that they sag a little under gravity.  
Any vertical gradient of the magnetic field therefore results 
               in a different average value of the magnetic field for the two species.
  This in turn  leads to a small shift in the mercury-corrected neutron Larmor frequency.  
For a given vertical gradient, this shift is in opposite directions for the two different orientations (up vs.\ down) of the main magnetic field.  
Furthermore, there is a systematic effect leading to a significant false-EDM  arising from a conjunction of the vertical magnetic-field gradient
  and the relativistic motional magnetic field seen by the mercury atoms (in particular)
  as they move through the electric field\cite{Pendleburry-GeometricPhase-2004}. 
We therefore needed to interpolate our measured nEDM results to zero vertical magnetic-field gradient.  
As we did not have an absolute gradiometer,
  we intentionally applied small magnetic-field gradients using trim coils in order to determine the situation at zero gradient
  from the intersection of the two curves arising from the two magnetic-field directions%
  \cite{Roccia-nEDM-data-taking-strategy-2018, Pendleburry-GeometricPhase-2004,PhysRevD.92.052008-Gravitational-depolarization}. 
 \TODOdone{cite for apparatus, co-magnetometer, R-curve}
 Hence, it is important to state that a blinding offset, in contrast to some kind of blinding factor, does not interfere with the interpolation of the curves.

\section{Data blinding}

\subsection{Blinding concept}
Any blinding method for an nEDM experiment must shift the measured Larmor frequency in correlation to the electric field, 
 ideally while leaving all other observables unaltered. 
The following blinding procedures were briefly considered by our collaboration:
\begin{itemize}
\item{}Apply a modified spin-flip frequency with respect to the recorded value during the experiment.
  However, this would modify the experiment in an insidious manner
    as the change in actual physical conditions applied would be correlated to the electric field changes. 
  This could therefore potentially introduce systematic effects, and,
    additionally, it would be irreversible. So, one would have no possibility to investigate it \textit{a posteriori}.
\item{}Register a shifted spin-flip frequency with respect to the one actually applied. This would require also subtle adjustments of all other magnetometer readings in order not to immediately reveal the shift by comparison. 
     In our case this would have meant to adjust consistently a total of 16 magnetometer readings (1 mercury and 15 caesium)%
     \cite{Roccia-nEDM-data-taking-strategy-2018,Cs-array} --- 
     A daunting task.
\end{itemize}
Both of these rejected techniques would mimic an EDM according to the usual analysis strategies.
There are other observables that can be modified, but they would not have exactly the same appearance as an EDM signal,
 and thus it would ultimately be fairly trivial to identify them as fake signals.
Note, that manipulating  the value of the electric field cannot be used to introduce a blinding offset.

The remaining variable that can be modified is the neutron counts.  
The primary difficulty in that case is 
 that, since the size of the required shift depends upon the neutron counts themselves,
 a partial but automatic analysis of the data must be done.
 And, in order to hand out as early as possible blinded data to the analysis teams,
 this must be done during data taking in a way that is fully defined before starting the actual data-taking campaign.
This was the approach that we have adopted for the nEDM experiment,
  and its implementation will be described in detail in the following sections.

\subsection{Algorithm}
The blinding algorithm operates stepwise.
First the necessary parameters are extracted from a full run (sect.\,\ref{sec:determination_of_alpha_and_detector_asymmetry}).
Then the position of each cycle on the Ramsey curve, the so-called working point, is determined (sect.\,\ref{sec-determining-working-point}),
 before the number of neutrons in each cycle can be modified (sect.\,\ref{sec-transfer-neutrons}).

\subsubsection{Calculation of the number of neutrons to be transferred}
\label{sec-calculation-neutrons-to-transfer}

\begin{figure}[btp]
  \centering
    \TODOdone{
      reduce vertical space and second $\nu$ axis between plots}
  \includegraphics[width=\jkfigurewidth]{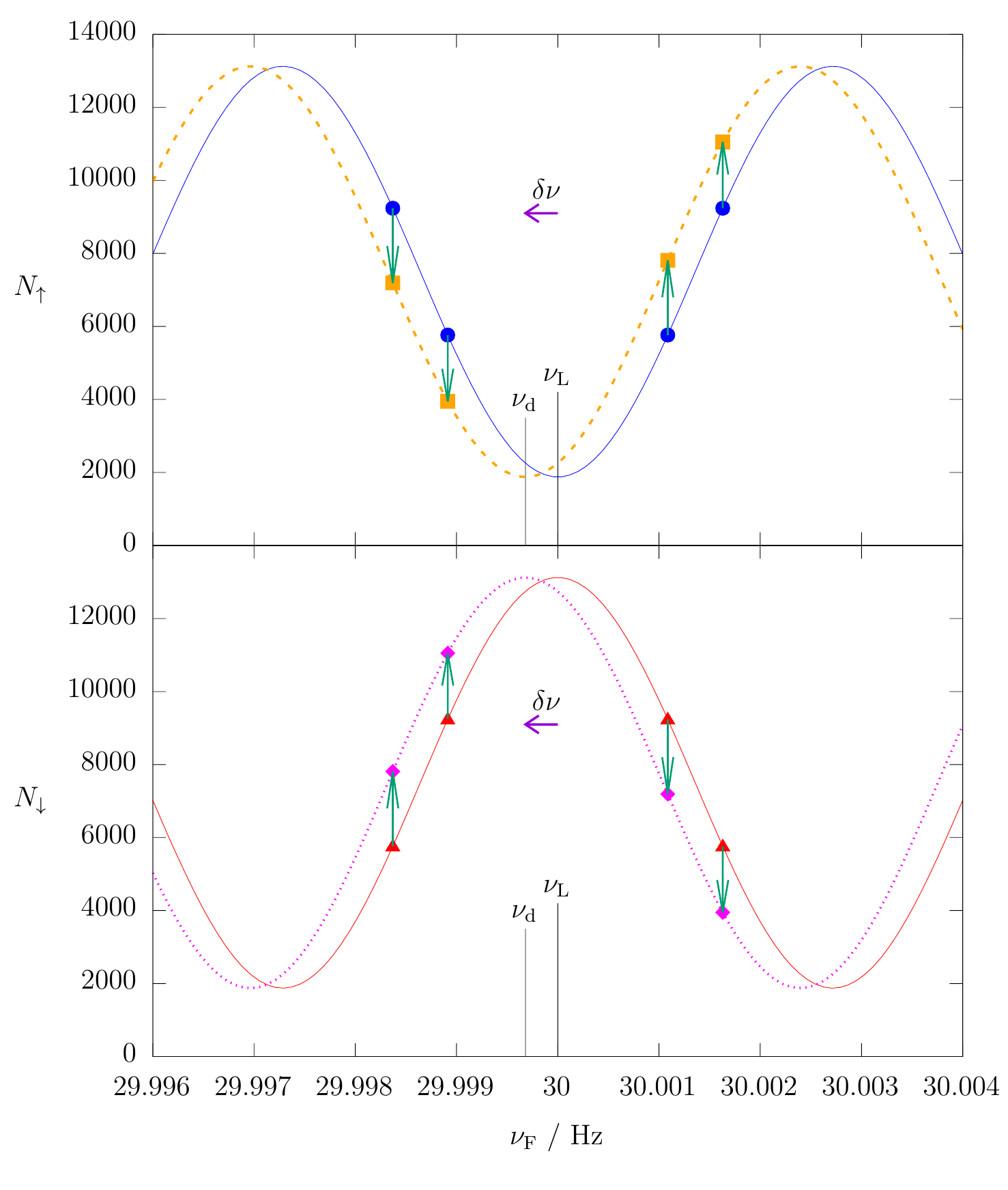}
  \caption{Simulated neutron counts plotted versus applied spin-flip frequency $\nu_\mathrm{F}$.
  	The transfer of a small number of neutrons (green arrows) from their initially recorded state,
  	         \eg counts $N_\uparrow$ (blue circles), corresponding to the original Larmor frequency $\nu_{\mathrm{L}}$,
  	         to the other spin state creates the blinded data points (orange squares).
    If this is done systematically and proportionally to the electric field, on can extract from the resulting dashed orange line a different Lamor frequency $\nu_\mathrm{d}=\nu_{\mathrm{L}}+\delta\nu$. 
    The frequency shift by $\delta\nu$ (violet arrow) represents a false EDM signal $d$ given by eq.\,\eqref{eq:delta_nu-resonanceshift-nedm}. 
    For the detector arm counting the opposite spin state, \eg $N_\downarrow$, the corresponding shift leads from the solid red to the dotted magenta curve. 
    This yields the same false EDM signal.
    In case SF1 is on, all points and lines must be mirrored at a horizontal line at $N=7500$.
    For clarity the strongly exaggerated values $\left|\sin\phi\right|=0.951$ and $d=\SI{3e-23}{\ecm}$ have been used here.}
  \label{fig:shift-ramsey-pattern}
\end{figure}

In order to generate an $\vec{E}$-field	dependent frequency shift
 a small number of spin-up neutrons have to be reclassified as spin-down, or \textit{vice versa}, see fig.\,\ref{fig:shift-ramsey-pattern}.

  We follow  eqs.\,\eqref{eq:phi_i}, \eqref{eq:detectedNeutrons-SpinUp}, \eqref{eq:detectedNeutrons-SpinDown} and \eqref{eq:delta_nu-resonanceshift-nedm} 
  as well as the first-order Taylor expansion
  $\delta N= \left( \frac{\mathrm{d}}{\mathrm{d}\phi} N \right) \left( \frac{\mathrm{d}}{\mathrm{d}\nu_\mathrm{L}} \phi \right)  \delta \nu$ to find the number of neutrons that need to change state:
\begin{eqnarray}
\delta N_{\uparrow;i} &=&   \epsilon_\mathrm{A} \frac{N_i^\prime}{2} 
\alpha^\prime \left( 2p_\mathrm{A} -1 \right)  \left( \sin\phi_i \right) \left( \frac{\mathrm{d}}{\mathrm{d}\nu_\mathrm{L}} \phi \right)  \delta \nu
\label{eq:numberOfNeutronsToBeShiftedNupOfDeltePhi}\\
&=&  -  \epsilon_\mathrm{A} \frac{N_i^\prime}{2} 
\alpha^\prime \left( 2p_\mathrm{A} -1 \right)  \left( \sin\phi_i \right)
\frac{\pi}{\Delta \nu}\frac{2 \, d \,   \vec{E} \cdot \vec{B} / \left|\vec{B}\right| }{h}
\label{eq:numberOfNeutronsToBeShiftedNup}\\
\delta N_{\downarrow;i} &=& +  \epsilon_\mathrm{B} \frac{N_i^\prime}{2} 
\alpha^\prime \left( 2p_\mathrm{B} -1 \right)  \left( \sin\phi_i \right)
\frac{\pi}{\Delta \nu}\frac{2 \, d \,   \vec{E} \cdot \vec{B} / \left|\vec{B}\right| }{h} \ .
\label{eq:numberOfNeutronsToBeShiftedNdown}
\end{eqnarray}
Note that $\vec{E}$ and $\vec{B}$ have to be parallel or antiparallel,
and that the sign of $\alpha^{\prime}$ can be negative depending on the state of SF1.

At this point, we make the convenient 
 approximation
that $\epsilon_\mathrm{A}=\epsilon_\mathrm{B}$,
which is reasonable in light of the performance of our detectors\cite{AfachHelaine-USSA,HelainePhD}.
This allows us to introduce the total number of measured counts per cycle  $N_i= \frac{ \epsilon_\mathrm{A} +\epsilon_\mathrm{B} }{2} N_i^\prime$ and its average over a run $\bar{N}=\,<\!N_i\!>$.
We assume further that the performance of our spin analysers is equal for the two spin states, that is $p_\mathrm{A}=p_\mathrm{B}$,
  which is, again, reasonable and supported by the data.
Thus, we can introduce $\alpha=\alpha^\prime(2 p_\mathrm{A} -1)$ which describes the measured visibility, but again, is negative in case SF1 is enabled.
Therefore,
\begin{equation}
\delta N_{\uparrow,\downarrow;i} =  \mp N_i
\frac{\pi\alpha}{\Delta \nu}\frac{d \, \vec{E} \cdot \vec{B} / \left|\vec{B}\right|}{h} \sin\phi_i \ .
\label{eq:ucns-to-be-shifted_Ni}
\end{equation}
We will discuss below the implications of removing the assumptions of $\epsilon_\mathrm{A}=\epsilon_\mathrm{B}$ and  $p_\mathrm{A}=p_\mathrm{B}$.

Typical values for the nEDM experiment are $N_i$ = \num{15000}, $\left|\sin\phi\right|$ = \num{0.99}, 
$\alpha$ = \num{0.75},
$T$ = \SI{180}{s}, $t$ = \SI{2}{s}, and $E$ = \SI{11}{kV/cm}.
Thus an EDM offset of \SI{1.0e-25}{\ecm} would require a shift of about 3.39 neutrons in each cycle.  
In terms of signs, bearing in mind that the neutron has a negative magnetic moment,
 if $\vec{B}$ and $\vec{E}$ are parallel, a positive nEDM
 would {\em reduce} the precession frequency.  
This would shift the Ramsey curves towards smaller frequencies,
 which means that neutrons measured at a working point above
 the resonant frequency
 shift from the spin-down detector arm to the spin-up.
 Neutrons that are measured below the working point shift from down to up, correspondingly.
Figure~\ref{fig:shift-ramsey-pattern} illustrates this reclassification and the resulting shift.

\label{text-introduce-fractional-neutrons}
Obviously, it is impossible to shift a non-integer number of neutrons in a single cycle.
One could simply round the number,
 but this would effectively cause a granularity of $\sim$ \SI{3e-26}{\ecm} in the available blinding offsets.
However, we can add to $\delta N$ a small random number with a normal distribution, and round the  sum to the nearest integer number. 
The choice of the width of this normal distribution was driven by two facts.
 On the one hand, a small width does not smooth the granularity sufficiently. 
On the other hand, a large width adds noise to the neutron counts and thus to the blinded nEDM value.
We found a suitable compromise using a standard deviation of 2 counts. 
In this case the granularity is sufficiently suppressed so that the result differs from a flat distribution by less than $10^{-7}$.
 \label{sec:non-integer-neutron-to-be-shifted}
An improved method will be suggested in sect.\,\ref{sect:improvement-dithering}.

As mentioned above, this algorithm assumes the same $\bar{N}$ and $\alpha$  for each of the two spin states.  
If this were not to be the case, a direct transfer of neutrons from one spin state to the other would not be appropriate.  
Instead, one would have to analyse and treat the two states separately, 
and neutrons would have to be added to or deleted from the spin-up and spin-down arrays as required.
While this is trivial if the neutron data merely consists of a simple sum of counts per cycle,
 it is a substantial effort for a more detailed data format such as ours, which lists charge and time per event.

\subsubsection{Determination of \texorpdfstring{$\alpha$}{alpha} and detector asymmetry} \label{sec:determination_of_alpha_and_detector_asymmetry}  
Before the data can be blinded one has to determine $\alpha$ and $\nu_\mathrm{L}$. 
While $\alpha$ is sufficiently constant throughout an entire run,
 $\nu_\mathrm{L}$ might change from cycle to cycle and must be corrected with the field values recorded by the mercury \comagnetometer.  
We therefore refer to it as $\nu_{\mathrm{L},i}$ and write 
\begin{equation}
	 \nu_{\mathrm{L},i} = \left|\frac{\gamma_\mathrm{n}}{\gamma_\mathrm{Hg}}\right| \nu_{\mathrm{Hg},i}
	      - \frac{\Phi}{\pi} \Delta \nu \ ,
	 \label{Lamor-frequency-vs-mercury-and-gradient-phase}
\end{equation}
where $\gamma_\mathrm{n}$ and $\gamma_\mathrm{Hg}$ are the gyromagnetic ratios of the neutron and mercury respectively,
  $\nu_{\mathrm{Hg},i}$  the frequency obtained from the mercury \comagnetometer,
  and   the phase $\Phi$ accommodates any difference in the average magnetic field sampled by the two species. 
We had measured the ratio of gyromagnetic ratios in a previous experiment\cite{Afach-gyromagnetic-ratio-n-Hg-2014}. 
For the blinding algorithm we used a fixed value of $\left|\frac{\gamma_\mathrm{n}}{\gamma_\mathrm{Hg}}\right| = 3.8424574$. 
As magnetic field gradients were not changed during a run, $\Phi$ kept the same value throughout all cycles of the run.

We rewrite eqs.\,\eqref{eq:detectedNeutrons-SpinUp} and \eqref{eq:detectedNeutrons-SpinDown} as
\begin{align}
   \frac{N_{\uparrow,i} - N_{\downarrow,i}}{N_{\uparrow,i} + N_{\downarrow,i}} 
     &= \frac{\epsilon_\mathrm{A}-\epsilon_\mathrm{B}-\alpha^\prime \left( \epsilon_\mathrm{A} (2 p_\mathrm{A} -1) + \epsilon_\mathrm{B} ( 2p_\mathrm{B} -1) \right) \cos \phi_i}
              { \epsilon_\mathrm{A}+\epsilon_\mathrm{B}+\alpha^\prime \left(-\epsilon_\mathrm{A} (2 p_\mathrm{A} -1) + \epsilon_\mathrm{B} ( 2p_\mathrm{B} -1) \right) \cos \phi_i} \\
     &\approx \frac{\epsilon_\mathrm{A}-\epsilon_\mathrm{B}}{\epsilon_\mathrm{A}+\epsilon_\mathrm{B}} - \alpha  \cos \phi_i  \ ,  \label{eq:asymmetry-simplified}
\end{align}
where we made the approximations  $\epsilon_\mathrm{A}\approx \epsilon_\mathrm{B}$ and $p_\mathrm{A} \approx p_\mathrm{B}$. Furthermore, we used that $\epsilon_\mathrm{A}-\epsilon_\mathrm{B} \ll   \epsilon_\mathrm{A} + \epsilon_\mathrm{B}$ which is also justified by the data.

We define $A_m=\frac{\epsilon_\mathrm{A}-\epsilon_\mathrm{B}}{\epsilon_\mathrm{A}+\epsilon_\mathrm{B}}$ which is the detector asymmetry
originating from the slightly different efficiencies of the two detector arms counting the two spin states.
We use eqs.\,\eqref{Lamor-frequency-vs-mercury-and-gradient-phase} and  \eqref{eq:phi_i} to rewrite eq.\,\eqref{eq:asymmetry-simplified} as
\begin{equation}
\frac{N_{\uparrow,i} - N_{\downarrow,i}}{N_{\uparrow,i} + N_{\downarrow,i}} 
= f\left( \nu_{\mathrm{F},i} - \left|\frac{\gamma_\mathrm{n}}{\gamma_\mathrm{Hg}}\right|  \nu_{\mathrm{Hg},i}    \right) \ ,
\label{eq:fit-equation}
\end{equation}
 where we have defined a function as:
\begin{equation}
	\label{eq:f_R}
	f(x) = A_m - \alpha \cos \left( \frac{\pi}{\Delta \nu} x + \Phi \right) \ .
\end{equation}
\TODOdone{JK: This is an approximation for large and roughly equal detector efficiencies and only for $\phi_i\approx
	 90^\circ$\\}
\TODOdone{JK; I did change to a - sign to match the definition \eqref{Ramsey-SpinUp} which is important for our SpinUp definition\\}
The independent variable $x$ is beneficial for the fit algorithm, since it can be calculated from the observables for each cycle. 
It represents the frequency difference between applied spin-flip and neutron resonance.
The parameter estimation of $A_m$, $\alpha$ and $\Phi$ is done by fitting the data of a full run to eq.\,\ref{eq:fit-equation}.
Every four cycles we inverted the spin-flipper states on both detector arms
  by activating and deactivating spin-flippers that are mounted inside the arms\cite{AfachHelaine-USSA,HelainePhD,Ban-NANOSC-2016}. 
This results in a ``Normal'' and an ``Inverted'' configuration, with asymmetries $A_{\mathrm{N}}$ and $A_{\mathrm{I}}$ respectively. 
Both values are almost constant throughout a run. 
We retained them as fit parameters to accommodate for long term changes. 
Consequently, the data contains two collated subsets and the fit must be conducted as a simultaneous fit, within which $\alpha$ and $\Phi$ are common parameters
 while $A_{\mathrm{N}}$ and $A_{\mathrm{I}}$ apply to the respective partial data sets only.

\subsubsection{Determination of the Ramsey phase \texorpdfstring{$\phi_i$}{phi\_i}}
\label{sec-determining-working-point}
After having carried out the fit on the full run
 we can  use eq.\,\eqref{eq:ucns-to-be-shifted_Ni} to calculate the number of neutrons to be transferred for each cycle. 
However, we still need to determine $\phi_i$. 
This may be done either via eq.\,\eqref{eq:asymmetry-simplified}:
\begin{equation}
 \cos \phi_i = \frac{1}{\alpha} \left( \frac{N_{\uparrow,i} - N_{\downarrow,i}}{N_{\uparrow,i} + N_{\downarrow,i}}  - A_m \right) \ ,
  		\label{eq:phi_i-via-neutrons} \\
\end{equation}
or via eq.\,\eqref{eq:phi_i}:
\begin{equation}
 \phi_i = \frac{ \nu_{\mathrm{F},i}- \left|\frac{\gamma_\mathrm{n}}{\gamma_\mathrm{Hg}}\right| \nu_{\mathrm{Hg},i} }{\Delta\nu}\pi +\Phi \ .
   \label{eq:phi_i-via-mercury} \\
\end{equation}
We have implemented the first variant, as it is more robust in instances where in a single cycle the \comagnetometer\ provides a bad reading.
Note that also this variant uses eq.\,\eqref{eq:phi_i-via-mercury} to determine the sign of $\phi_i$.

\subsubsection{Transferring neutrons}
\label{sec-transfer-neutrons}
The data files are an event-driven list where each entry consists of 
 a time stamp, the integrated charge recorded at the time, 
 and the identification number of the photomultiplier tube (PMT) that observed the event\cite{Ban-NANOSC-2016};  
if the charge exceeds a certain threshold then the event is classified as a neutron detection.
Each of the two detector arms, one per spin state, consists of a set of nine PMTs, which are sequentially numbered from 0 to 17 with 0--8 in the first arm and 9--17 in the second.
In order to reclassify the spin of a neutron it is therefore sufficient to take the PMT number of that event, add 9 and do a modulo 18 operation.
A neutron that is to be transferred is chosen by randomly selecting an event from the list counted in the correct detector arm,
 and then checking whether it is suitable to be moved simply and cleanly across:
 the requirement is that there must be a minimum separation in time between the event in question and the previous and subsequent events. 
We apply this condition to both the source and the recipient channel. 
The reason is to avoid the transfer of events for which the charge is split between neighbouring PMTs,
 or where the baseline correction algorithm has or would have to modify the charge\cite{faster}.
  \TODOdone{Have I understood this properly?? - PH. JK: No, I rephrased it. Is it now clear?}%
 If the event is not suitable, another randomly chosen event is tested until an appropriate one is found.

\subsection{Choice of the blinding offset}
\begin{figure}[bt]
	\centering
	\includegraphics[width=\jkfigurewidth]{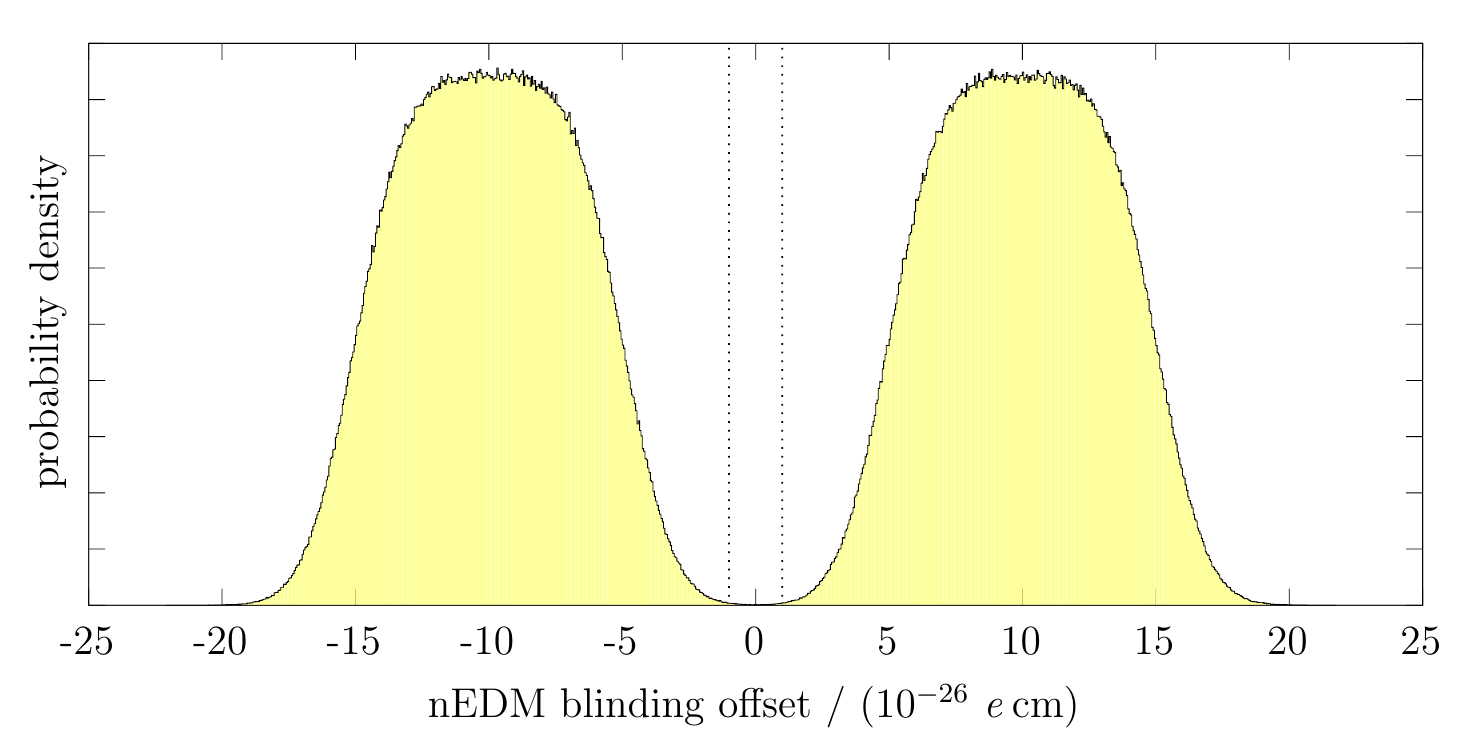}
	\caption{Probability density function for the choice of the blinding offset created with $10^6$ samples.
		The dashed vertical lines indicate the $\pm1\sigma$ sensitivity of the data accumulated in 2015 and 2016 assuming a mean value of 0. 
		For psychological reasons we keep the probability that an offset in this range is selected very small but non-zero (integrated probability $\approx 2\times10^{-4}$).
	}
	\label{fig-blinding-offset-distribution}
\end{figure}
Obviously, the value of the blinding offset must be kept secret from the analysis teams. 
In order not to provide any indirect psychological bias as to what it might be it should be randomly chosen from a distribution that allows a wide range of values.  
It is convenient for its modulus to be larger than the known upper limit of the nEDM,
  since this allows a ``sanity check'' of having a sign that can be confirmed for consistency prior to publication of results
  (see sect.\,\ref{sec:cost-and-benefit}).

At the same time, it should be sufficiently small in order to guarantee that the working points are not shifted away from the steep slope of the Ramsey 
pattern %
such that we maintain the sensitivity and the Taylor expansion used in eqs.\,\eqref{eq:numberOfNeutronsToBeShiftedNupOfDeltePhi}--\eqref{eq:numberOfNeutronsToBeShiftedNdown} holds.
Any error in the calculation of the number of neutrons that are shifted by the blinding process 
 will add noise to the EDM signal and therefore make it more difficult
 to look for effects and correlations that might indicate possible systematic effects such as the motional-field effect described above.   
\TODOobsolete{how much are the workingpoints shifted by this - WP have an neutron asymmetry of +- 880, thus 3/880}
\TODOobsolete{can one attack the blinding through this shift? - If AcqEDM would fit perfectly - No, because neutron transferring is exactly good for this reason.}

For the nEDM experiment we chose a combination of four Heaviside step functions
 that together define a range of \SI{\pm 15e-26}{\ecm}
 and exclude a modulus smaller than \SI{5e-26}{\ecm}. 
We then blurred this function with a Gaussian of width \SI{\pm 1.5e-26}{\ecm}.
We also explicitly excluded the extremely unlikely possibility that the tail of the Gaussian would extend beyond \SI{\pm 1e-24}{\ecm},
 in order to ensure that we stay within the linear region of the original Ramsey fringe.
One could argue that this  latter step represents a small psychological bias,
 but --- notwithstanding the previously existing world limit ---  
 a one-day measurement without blinding leads to the certain conclusion
 that the true nEDM value must be smaller.
Finally, we also excluded  a modulus of \SI{< 1e-28}{\ecm} for technical reasons,
since when communicating  between different programs we use a value of exactly zero for cycles that should not be blinded at all,
 \eg those with no applied electric field. 
Figure~\ref{fig-blinding-offset-distribution} shows the probability distribution of the blinding offset.
 
\subsection{Secondary blinding and reblinding}

\begin{figure}
	\centering
	\begin{tikzpicture}[scale=\myTikzStrechFactor, every node/.style={scale=0.7}]
	  \node (rawdata)          at ( 0.5,0) [fill=gray!30,draw,align=center] {Raw\\data};
	  \node (offset)           at ( 0.5,-2) [fill=gray!30,draw,align=center,rounded corners] {Offset\\primary blinding};
	  \node (blinding_primary) at ( 3,0) [fill=gray!30,draw,align=center,ellipse] {Blinding};
	  \node (data_primary)     at ( 6,0) [fill=gray!30,draw,align=center] {Blinded\\data\\(primary)};
	  \node (offset_eastern)   at ( 6,+2) [fill=gray!30,draw,align=center,rounded corners] {Eastern offset\\secondary blinding};
	  \node (offset_western)   at ( 6,-2) [fill=gray!30,draw,align=center,rounded corners] {Western offset\\secondary blinding};
	  \node (blinding_eastern) at ( 8.5,+1) [fill=gray!30,draw,align=center,ellipse] {Blinding};
	  \node (blinding_western) at ( 8.5,-1) [fill=gray!30,draw,align=center,ellipse] {Blinding};
	  \node (data_eastern)     at (11.5,+1) [fill=gray!30,draw,align=center] {Blinded data\\(eastern)};
      \node (data_western)     at (11.5,-1) [fill=gray!30,draw,align=center] {Blinded data\\(western)};
	  
	  \draw[->, blue!50, very thick] (rawdata)          to                (blinding_primary);
	  \draw[->, blue!50, very thick] (offset)           to[out=0, in=-90] (blinding_primary);
	  \draw[->, blue!50, very thick] (blinding_primary) to                (data_primary);
	  \draw[->, blue!50, very thick] (data_primary)     to[out=0, in=-130] (blinding_eastern);
	  \draw[->, blue!50, very thick] (data_primary)     to[out=0, in=130]  (blinding_western);
	  \draw[->, blue!50, very thick] (offset_eastern)   to[out=0, in=90] (blinding_eastern);
  	  \draw[->, blue!50, very thick] (offset_western)   to[out=0, in=-90] (blinding_western);	  
	  \draw[->, blue!50, very thick] (blinding_eastern) to                (data_eastern);	  
	  \draw[->, blue!50, very thick] (blinding_western) to                (data_western);		  
	\end{tikzpicture}
	\caption{Illustration of primary and secondary blinding. 
		    Each analysis group has access only to their respective blinded data set, ``eastern'' or ``western''.}
	\label{fig:secondary-blinding}
\end{figure}
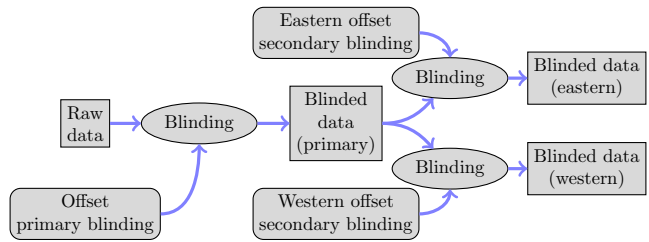
The nEDM collaboration decided before data taking to have the analysis carried out by two independent teams,
 referred to as eastern and western,
 loosely reflecting the geographic distributions of the involved institutions.
In order to allow them to communicate without introducing a bias in case of any discrepancy over the mean value of the nEDM,
  it was decided that in addition to the first-stage ``primary'' blinding
  we would use exactly the same algorithm to apply a separate ``secondary'' blinding that was distinct for each group,
   \ie with a different offset.  
Figure~\ref{fig:secondary-blinding} illustrates this process. 
  
During the early days of data taking some concern was expressed that the automatic fitting algorithm might not work properly in all cases,
 or that some important properties of the data might be hidden as a result of the blinding,
 or that some other similarly unexpected events might make it necessary to significantly change the blinding algorithm.  
In order to provide a consistent data set in any of those cases,
 it would be necessary to run a modified blinding program again from scratch on the raw data.  
However, since the first set of blinded data would by then already be available to the analysis teams,
 it would be trivial for them  to compare two versions of the same data file,
  and by leaving out all mismatching events they would have an unblinded data set with a statistical significance close to the original data set.  
In order to avoid such a scenario we made sure that our pseudo-random number generator delivers reproducible numbers,
 and that the neutrons that are transferred are selected reproducibly.
Thus, if \eg one version of the blinding algorithm shifts seven neutrons and the other eight within a given cycle,
 the two resulting files would only differ by one neutron for that cycle.  
Therefore, a reblinding using the same or similar offset and a slightly modified algorithm can be carried out without danger of inadvertent unblinding. 
It shall be mentioned that reblinding with an offset of opposite sign would immediately reveal both offsets.

In addition to transferring the neutrons between spin states, 
 the blinding algorithm also marks each blinded data file with the date of blinding and the version number of the blinding code in order
 to ensure that those otherwise very similar files remain clearly distinguishable.

After all, the reblinding feature was not used, since in our data analysis no large discrepancies occurred.

\subsection{Pseudo-random number generator (PRNG)}
In principle, the  random numbers used should meet the same strict requirements
 as those for strong cryptography regarding the prediction of numbers,
 the correlation between them and the uniformity of their distribution. 
However, the quantity of random numbers that we need is very small --- typically a dozen per cycle for about 50000 cycles, where each cycle gets a new seed. 
Therefore, a prediction attack to reveal the blinding offset would be extremely unlikely to succeed even if the random number generator is not of highest quality.
In contrast, the quality of the generator is important in terms of non-correlation and uniformity
 in order to avoid the danger of introducing noise or a systematic bias to the blinded data.
The standard PRNG of many computer languages, the linear congruential generator, therefore may be not suitable.
Furthermore, for the reblinding  it is absolutely necessary that the same algorithm should remain available  for a significant number of years.  
Thus, we avoided any sort of library that may vary either over time or between different computers.  
We decided to use WELL1024a\cite{Panneton:2006:ILG:1132973.1132974}. %
The Box-Muller transform\cite{box1958} was used to convert uniformly distributed to normally distributed random numbers where necessary.

The random seed for each cycle must be reproducible over years, and it must be secret after  blinding. 
Our data format, which, as noted above, is an event list of particle detection per channel,
 includes a periodic counter of accumulated events in every channel.
 This led to the choice of using a 1024-bit checksum over the last \SI{130}{kByte} of the unblinded file. 
Note that if the data files were not to include such a counter,
 the blinded data file would be very similar to the unblinded one and the seed would not be secret.
In such cases an alternative approach would be to use the noise in the detector for the seed creation, \eg from gamma events.
For the secondary blinding, the original unblinded data were used for the seed calculation. 
This would help if a reblinding at both primary and secondary level were ever to become necessary.
\TODOdone{reorder the sentences above}

\subsection{Online blinding}
In order to calculate the phase of the actual working point $\phi_i$ the blinding process requires the knowledge of $\Phi$ and $\alpha$.
Yet, this information is available only after a full run has been recorded, since it results from the overall Ramsey fit. 
Consequently, no blinded data are available before the end of each run.
However, it is absolutely necessary to have some live data available for quality checks of the ongoing measurement.
An intuitive thought would be to publish a rounded version of the neutron counts in order to disguise the blinding offset.
 It turns out though, that, in order to make this disguise effective,
 the rounding must be so coarse
 that the obtained number would be useless as a quality check.
As a solution to this problem, we devised an online blinding mechanism. 
For this, an additional blinding offset was created randomly for each run. 
The range for these random numbers with uniform probability distribution was \SI{\pm 1e-23}{\ecm}
 and thus about a hundred times larger than the range of the regular blinding offset.
The list of online blinding offsets used was stored in a location with restricted access, and has  not been used for any other purpose than debugging the program.
The online blinding algorithm does not provide a $\Phi$ parameter, that means it  assumes a zero magnetic field gradient for the calculation of the number of neutrons to be shifted.
Furthermore, it assumed perfectly symmetric detector efficiencies and uses $\bar{N}$ instead of  $N_i$.
In all other aspects the algorithm was identical to that used for the regular blinding.  
The frequency shift introduced through the online blinding was of the same order of magnitude as the naturally occurring magnetic field gradients,
either intentionally introduced by the trim coils, or by external fluctuations.
With the online blinding system in place, it was possible to make neutron counts available to the user immediately after each cycle
 without affecting the quality of the data, as far as online checks are concerned.  
However, these data can obviously not be used meaningfully for any further analysis.

It is important that cycles without electric field are not blinded at all,
 since they are used by the DAQ to choose the working points to have symmetric neutron count rates even in the presence of magnetic field gradients.

\subsection{Technical details}
The supervisory control and data acquisition system of the experiment was partly modular, with
all processes and file handling concerning neutron counting being hosted on a dedicated computer running Linux. 
Time was synchronised among all computers, and control communication was done via Ethernet (TCP/IP). 
Thus, with simple user permissions provided by the file system we were able to restrict access to the binary code of the blinding program,
 which contains the blinding offset, as well as to the raw data files.  
It was particularly beneficial that the computer could be started with a common unprivileged account.  
The DAQ program and thus the blinding process were given different permissions via the setuid bit.
Consequently, the blinding process had access to the secret blinding offset and could write data files that standard users could not read.

A typical run of several days generated about a dozen gigabytes of data. 
With files of this size, the blinding process took several minutes.
We obviously desired immediate feedback about the blinding and any potential problems, but we did not wish to block our DAQ system for as long as that.
We therefore split the blinding process into two parts.
The first part was to select data and to do the fit of $\alpha$ which was reported to the main DAQ and thus to the user.
This could be carried out within a second. 
The process would then fork itself, on the one hand quitting to make the system available for the following run,
 while, on the other hand, simultaneously carrying out the intensive work of transferring neutron data between the two detector arms. 

During data taking the blinding program was supposed to run autonomously and without intervention. 
This meant it had to handle some irregular conditions:
\begin{itemize}
	\item{}Data that did not contain EDM information must not be blinded; they were instead revealed immediately. 
	These were typically runs without applied high voltage, or runs with cycles that did not have two spin-flips. 
	Such measurements occurred fairly frequently in order to characterise the UCN source\cite{PSI-Source-Characterization-EPJA-2020,Lauss2014}, the detector, or the background.
	\item{} The fitting process ignored single cycles with a low neutron count rate. 
	        We chose 1000 counts as a threshold, as such a low count rate would not be used for nEDM analysis.
	\item{} Cycles with an unphysically  high count rate were not blinded, since these could effectively disclose the blinding offset. 
	        We chose a threshold of 50000 neutrons per cycle, 
	        ten times more than during commissioning and nearly three times more than the maximum observed.
	\item{} Blinding was only automatically applied if the quality of the Ramsey fit was sufficiently good, we chose  $\chi_\text{red.}^2 < 3$. 
\end{itemize}
In case of doubt the blinding process neither blinded nor revealed data, but rather made a request via E-mail for human intervention.
\TODOdrop{check code for other exceptions the code does handle}

\subsubsection{Manual interventions}\label{sec:manual_interventions}

We took great care during the design of the blinding algorithm to minimize the need for human intervention during data taking. 
This required automatic handling of unusual circumstances with
respect to data quality or malfunctions of parts of the apparatus. 
Inevitably, due to
the complexity of the experiment, some manual interventions during data taking were necessary. 
In these circumstances the data were assessed
by the blinding coordinator in order to either reject bad cycles, and to
apply the blinding on the remaining cycles of the run, 
 or divide a run into pieces between magnetic field jumps and apply the blinding on these parts.
During data taking in 2015 and 2016, 1072 runs with data files from the neutron detector were recorded.
Of these, 113 runs were automatically blinded, 14 runs needed manual blinding 
and 20 runs needed manual revealing. 
All of these contained information on the EDM.
The other 925 runs were revealed promptly. 
This is important as these were special characterization measurements that were required immediately for detailed analysis.

\TODOdone{move the information above to other chapters  (Manuel interventions + unblinding), then  cancel ``Chronology''}

\subsubsection{Secrecy} %
The scenarios we want to protect against are the following:
\begin{itemize}
	\item During the data analysis process somebody might be inclined to do a test that would be simpler to run on the unblinded raw data. 
	\item If forbidden data exists some human beings are tempted to try accessing them, only to prove that they can.
	\item Others may seek the challenge that there is something which is claimed to be impossible, namely to decrypt the data or to apply statistical attacks on them.
\end{itemize}
Any of them might be without malicious intent, but it may lead to accidental display of the blinding secret which is a simple number.

The blinding offset was stored
using asymmetric encryption with the public part of a 
RSA-key pair directly after it had been created.
The blinding offset together with some metadata only amounts to 192 bits, thus a simple asymmetric encryption is possible.
The private key to decrypt the blinding offset was injected into
the executable of the blinding program at compile time. 
Access to the executable program is restricted by file system permissions.
The original private key was stored with password RSA encryption using OpenSSL and thus only available to the blinding coordinator.
Access to data files was restricted by file system permissions.

These cryptographic and organisational measures were deemed reasonable  in order to prevent accidental unblinding of the data. 
They were easy to implement and did not impact any permitted workflow.
Although fairly robust, they are certainly not
sufficient to protect against either physical theft of hard drives
or manipulation of software with malicious intent.
Any further protection would require the restriction of physical access to the DAQ computer or its boot process. 
Encryption of the operation system via a Trusted Platform Module chip is nowadays available and would suffice for this task.
However, this would have been a potential impact on the maintainability of the system, especially in case of hardware problems,
 hence we considered the existing hurdles high enough for our case.

\subsection{Effects of noise and asymmetry} \label{sec:noise}
The blinding algorithm manipulates the data. This includes the use of random numbers and fit results.
This necessary procedure naturally introduces some noise. 
In this section we discuss the level of this noise and the resulting consequences.

\subsubsection{Noise from fractional neutron numbers}
\label{sec-noise-from-fractional-neutron-numbers}

In sect.\,\ref{sec:non-integer-neutron-to-be-shifted} we described how a random number (normal distribution with $\sigma=2$) is added to the fractional number of neutrons to be transferred before rounding to an integer value.
Solving eq.\,\eqref{eq:ucns-to-be-shifted_Ni} for $d$ allows one to calculate how much noise is added to the final nEDM result due to this additional random process.
Using the average number of neutrons per cycle $\overline{N} = 11400$, the average visibility $\overline{\left|\alpha\right|} = 0.75$, and the applied electric field $E=\SI{11}{kV/cm}$
 the additional noise amounts to \SI{7.7e-26}{\ecm} per cycle. 
The additional statistical uncertainty for the mean of all 54068 cycles is \SI{3.3e-28}{\ecm},
  which is about 3\% of the uncertainty due to counting statistics.

\subsubsection{Noise from detector asymmetry}
\label{sec-noise-from-detector-asymmetry}

In sect.\,\ref{sec:determination_of_alpha_and_detector_asymmetry} we described the determination of $\alpha$ and $A_m$ through fitting.
These quantities each have their own statistical uncertainty.
The mean of the fit value of the visibility $\left|\alpha\right|$ was 0.75, and the mean of its uncertainty was 0.003.
The mean values of the detector asymmetry were $A_{\mathrm{N}}=0.032$ and $A_{\mathrm{I}}=-0.036$ in 2015; both with a standard deviation of 0.002.
 In 2016 the mean was $\left|A_m\right|=0.004$ with a standard deviation of 0.001.
 The mean of the individual uncertainties within each run was always below 0.001. 
 Thus in 2015 there was a significant asymmetry.
The number of neutrons to be transferred is calculated from these numbers via eqs.\,\eqref{eq:phi_i-via-neutrons} and \eqref{eq:ucns-to-be-shifted_Ni}.  
At our working points the result of $\sin(\arccos(x))$ lies between 0.98 and 0.99 for any $x$. 
Thus no matter how large the fluctuations of $A_m$ may have been,
 the resulting noise on $\delta N$ is less than 1\% and thus much smaller than the noise arising from the integer rounding described in sect.\,\ref{sec-noise-from-fractional-neutron-numbers},
 and is thus also negligible.

\subsubsection{Noise from visibility}
\label{sec-noise-from-visibility}
  
The parameter  $\alpha$ enters directly in eq.\,\eqref{eq:ucns-to-be-shifted_Ni},
 but as the observed relative uncertainty $\frac{0.003}{0.74}=0.004$ is also very small, the same argument applies once again.

\TODOdone{In eq.\,\eqref{eq:ucns-to-be-shifted_Ni} also the measured quantity $N_i$ does enter. 
	If this is considered as an observable this would have a large scatter ($\sqrt{15000}$), 
	However, JK thinks it does not add any noise, as it is the \emph{exact} number for this cycle. 
	??TODO Somebody should verify this.}  %

\subsubsection{Noise from neutron number per cycle}
\label{sec-noise-from-neutron-numbers}

In eq.\,\eqref{eq:ucns-to-be-shifted_Ni} also the measured quantity $N_i$ enters. 
Despite being a noisy observable, it does not contribute to any noise in the blinding,
 since it is the exact value of the number of neutrons for this particular cycle.

\TODOdrop{JK: probably leave away: our fit does not have an edm parameter,
	  If there were to be a real EDM of significant size,
	   then runs with different numbers of positive vs.\ negative HV cycles would have a bad determination of $\nu_\mathrm{L}$.  
	  In principle this could contribute to  noise, and potentially even to a bias; but in reality any such effects would be extremely small.}%

\subsubsection{Verification on test data}
\label{sec-noise-verification-on-test-data}

For a test of the blinding process with real data,
 we took advantage of the very first data taken before September 13th in 2015. 
This part of the data was made available to the analysis teams with and without a blinding offset, though the blinded one was not used in the analysis presented in ref.\cite{Abel2020PRL}.
It was blinded with $d^*=\SI{+1.951e-25}{\ecm}$ to test the blinding algorithm.
These 24 runs have each an irreducible statistical sensitivity that ranges from \SI{0.9e-25}{\ecm} to \SI{2.4e-25}{\ecm}. They accumulate to \SI{3.2e-26}{\ecm}.
The data was analysed in two stages, once in September 2015, right before the decision was taken to continue with the full implementation of the blinding,
 at this time obviously with a still quite rudimentary data analysis.
A second time this test was done with an almost final %
analysis. Both tests showed that the blinding algorithm increased the uncertainty by \SI{2e-28}{\ecm} corresponding to 0.5\% of the statistical sensitivity of the data set.
The blinding offset predicted by the  analysis matched the applied within \SI{0.2e-26}{\ecm} which was a tenth of the uncertainty of the analysis. 
This comparison was done before any unblinding in order to have a better estimate which outcome of the secondary (relative) unblinding should be considered as successful.
After unblinding, this test was repeated with the full dataset as described in the next section.
  		
\section{Unblinding}
Each data analysis team worked on a twice (primary + secondary) blinded data set, and  ultimately extracted their own estimator for the blinded nEDM value and its uncertainty. 
Once the collaboration was convinced that these analyses were complete, 
a comparison based on appropriate parameters and distributions was done. 
One comparator was for example the nEDM uncertainty. 
Moreover, since the data taking was organized in sequences,
  it was possible to check, sequence by sequence,
  the difference between the extracted nEDM and its mean value (averaged over all sequences). 
This difference was useful to check that the two analysis results showed the same correlations with respect to external parameters.
 
The decision to proceed to the first un-blinding step, which consisted of removing the secondary blinding offsets, was taken based on the agreement of all comparators. 

After this first unblinding it was possible to cross-check the two analyses with respect to the secondary blinding offset, see table~\ref{tab:TableUnblinding}.

\begin{table}%
	\centering
	\begin{tabular}{l|c|c}
		\hline
		\rule{0pt}{3.2ex}nEDM estimator & Western analysis team & Eastern analysis team  \\ \hline 
		\rule{0pt}{3ex}Twice blinded $\doubletilde{d}$         & \!\!\SI{15.39e-26}{\ecm} & \SI{3.80e-26}{\ecm} \\
		               Single blinded $\tsup[1]{d}$             & \SI{5.97e-26}{\ecm} & \SI{6.15e-26}{\ecm} \\
		                    non-blinded $d$ & \!\!\!\!\SI{-0.02e-26}{\ecm} & \SI{0.16e-26}{\ecm} \\
		\hline
		\rule{0pt}{3ex}$\doubletilde{d} - \tsup[1]{d} $         & \SI{9.55e-26}{\ecm} & \!\!\!\!\SI{-2.24e-26}{\ecm} \\
		               Input offset $d^{\prime\prime}$          & \SI{9.48e-26}{\ecm} & \!\!\!\!\SI{-2.33e-26}{\ecm} \\
		\rule{0pt}{3ex}Difference $\doubletilde{d} - \tsup[1]{d} -d^{\prime\prime}$ 
		                                                        & \!\!\!\!\SI{-0.05e-26}{\ecm} & \!\!\!\!\SI{-0.02e-26}{\ecm} \\
		\hline
		\rule{0pt}{3ex}$\tsup[1]{d} - d $                        & \SI{5.99e-26}{\ecm} & \SI{5.99e-26}{\ecm} \\
		               Input offset $d^{\prime}$                 & \SI{6.02e-26}{\ecm} & \SI{6.02e-26}{\ecm} \\
		\rule{0pt}{3ex}Difference $\tsup[1]{d} - d -d^{\prime}$  & \!\!\!\!\SI{-0.03e-26}{\ecm} & \!\!\!\!\SI{-0.03e-26}{\ecm} \\
		\hline
	\end{tabular}
	\caption{Estimators of the neutron EDM by the two analysis teams. $\tsup[2]{d}$ is the estimator of the twice blinded data. ${\tsup[1]{d}}$ is the estimator of the primarily blinded data. The input offset $d^{\prime\prime}$ is the value of the secondary blinding offset which was de-encrypted during the first, relative unblinding on October 23rd 2019. The input offset $d^{\prime}$ is the value of the primary blinding offset which was de-encrypted during the second unblinding on November 28th 2019.	All analysis results in this table comprise only data taken after September 13th 2019, data prior to this was not blinded with the same offsets and thus cannot be compared. Consequently, the value $d$ listed here differs slightly from the final result\cite{Abel2020PRL}.   
	The statistical uncertainty of the never-blinded data shown here was determined by one analysis team to be  \SI{1.07e-26}{\ecm} with $\chi^2/N_{\mathrm{dof}}=92.5/86$.
    }
	\label{tab:TableUnblinding}
\end{table}

Thereafter we compared directly the nEDM values obtained by the two teams. 
If any discrepancy would have been found,
 a longer and detailed comparison would have had to be carried out at this point.  
Among other things, we have discussed to run both analysis codes on a common subset of data and converge parameters and code, \ie cut criteria and methods, until the results match.
Another option would have been to produce a new set of secondary blinded data,
 although this would have been of limited use since by then both analysis teams would implicitly know their offsets.  
For this reason we rather had considered to produce an alternative blinded data set directly from the original raw data with a new unknown random offset.

Since the two analysis teams were in agreement   
 we proceeded to the removal of the primary blinding after the full completion of the evaluation of all systematic effects. 
The offset itself was revealed and subtracted, to yield a true nEDM estimator.  
In addition though,  the same analysis codes together with the same settings, \eg for cuts, were  applied to the original, non-blinded data set, which had been kept hidden up to that point.  
The result of the direct analysis of this never-blinded data set should match with that emerging from the analysis of the blinded set minus the applied blinding offset. 
From theoretical estimation as well as from our experience with the early data taken without blinding we expected this agreement to be on the $10^{-27}$\,\ecm\ level. In the posterior comparison, given in table~\ref{tab:TableUnblinding}, this was perfectly confirmed. 
\begin{figure}[bt]
	\centering
	\includegraphics[width=1.3\jkfigurewidth]{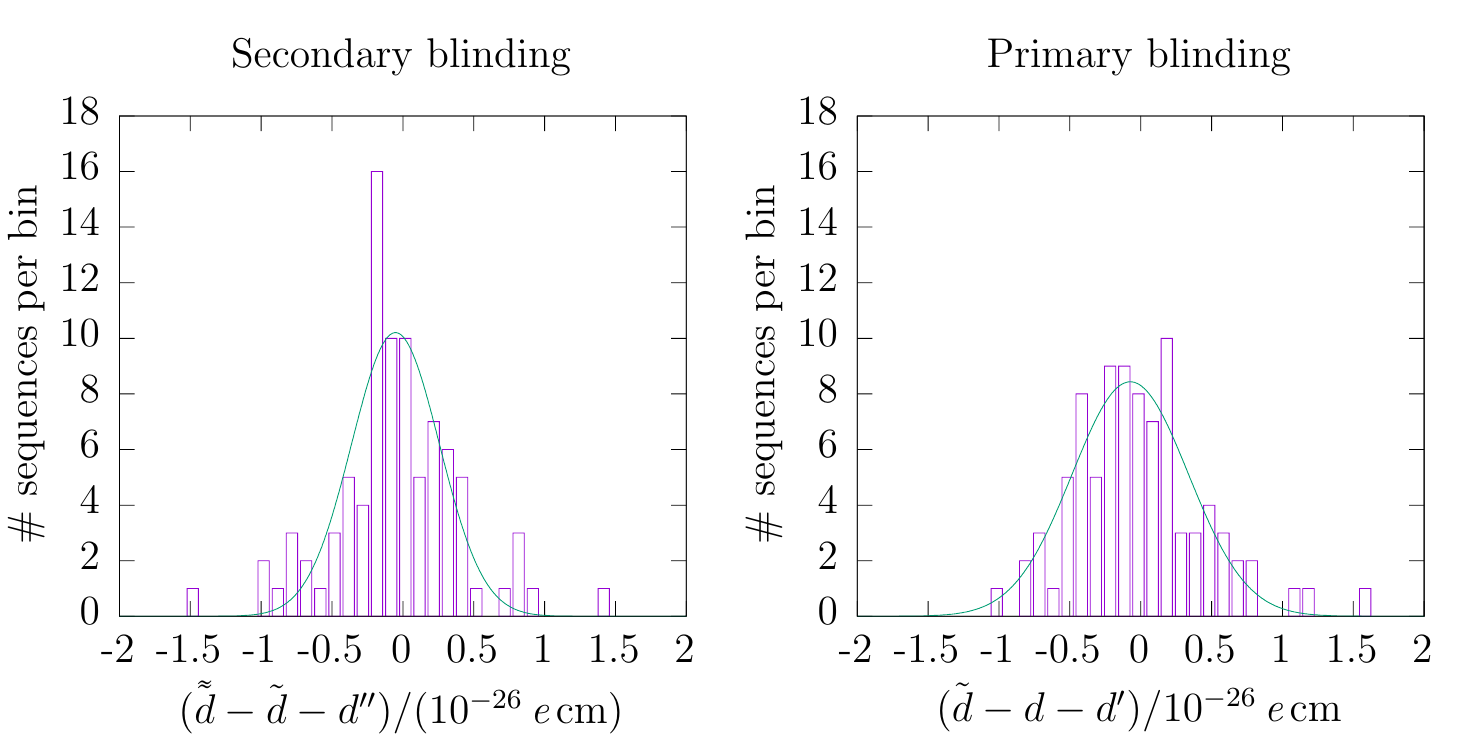}
	\caption{Difference of analysis on blinded and unblinded data sets and the corresponding offset for the two blinding steps.
		 The bin width is \SI{e-27}{\ecm}. 
		 Both peaks are centred well within \SI{e-27}{\ecm}. 
		Only data of the western analysis and after September 13th 2015 is shown. 
		The eastern analysis yields similar results.
	}
	\label{fig-unblinding-comparison}
\end{figure}
Figure~\ref{fig-unblinding-comparison} shows a comparison between the injected blinding offset and the one predicted by the analysis from the western team.
The non-zero width of the peaks indicates that the blinding algorithm does inject some noise into individual sequences or cycles.
The widths of the Gaussian fitted to the distribution were \SI{0.31(5)e-26}{\ecm} and \SI{0.41(4)e-26}{\ecm} for secondary and primary blinding, respectively.
And indeed, the blinding injects some noise in each cycle as explained in sect.\,\ref{sec-noise-from-fractional-neutron-numbers}.
The sequences consisted of 514 cycles on average.
Thus, the observed widths are compatible with the expected uncertainty of the mean due to the noise of \SI{0.34e-26}{\ecm}.

The agreement of the difference `analysis of blinded data' minus `analysis of never-blinded data' with the blinded offset is marvellous.

\TODOobsolete{Check whether this phrase is really escape-proof}

\section{Costs and benefits}\label{sec:cost-and-benefit}
As discussed in ref.\cite{Golub-AntiBlinding2014}, blinding does not come for free.
For the method presented here, the costs were primarily the manpower required for design, implementation and study of the technique. 
As noted above, our blinding introduces a small amount of statistical noise into the blinded data.
This tiny, additional noise was only present in the blinded datasets; the final analysis was never affected by this. 
Ultimately, we did not suffer from various costs that are typical for other blinding techniques. 
For example, in our case all analysis channels were immediately available and no signals or features, other than the true nEDM itself, was hidden.
Notably, the blinding permitted the analysis of a periodically changing nEDM\cite{Abel-2017-RawlikAyres-OscilatingEDM-AxionlikeDarkMatter} without revealing the unblinded result of the static signal.   

Most importantly, the blinding provided a very substantial benefit to the nEDM analysis, not only in that it eliminated the effects of an unconscious bias.   
Since in the past the true nEDM result has always been indistinguishable from zero, it was sign insensitive,  and as such also insensitive to potential sign errors in the analysis. 
However, once blinded, the signal in the data had a value significantly away from zero, and thus included a clearly identifiable sign.   
This sign showed up in the various analysis channels, \eg with its dependence upon magnetic-field gradients, and as such at one point it actually revealed a mistake in an early version of our data analysis when the code was tested with a known fake nEDM.

\section{Possible Improvements}
\label{sect:improvement-dithering}
In order to handle the non-integer number of neutrons to be transferred in each single cycle
we used a normally distributed random number of width~2 as described in sect.\,\ref{sec:non-integer-neutron-to-be-shifted}.
Future implementations will use a rectangular probability density function of width~1. 
This will provide perfect linearity and will reduce the introduced noise by a factor of two, down to the intrinsic minimum.

\section{Summary \& Conclusion}
We have developed and applied for the first time for an nEDM measurement a blinding technique.
Our algorithm only modifies a copy of already recorded data and saves the original data on a hidden and secret disk.
Secondary blinding and the re-blinding option are innovations which reduce risks that are often associated with blinding.
The artificial increase in noise in the blinded datasets through the process was negligibly small and disappears automatically in the final result.

\section{Acknowledgement}
We gratefully
acknowledge financial support from the Swiss National
Science Foundation through project grants
149211, 162574 
and 172639 for the ETHZ group, %
grant 181996 for the Bern group %
and grants 
 117696, %
144473, %
137664, %
163413, %
169596, and %
171626 %
for the PSI group.
This work was funded in part by the United Kingdom Science and Technology Facilities Council (STFC) through grants ST/N000307/1 and ST/M503836/1,
 as well as by the School of Mathematical and Physical Sciences at the University of Sussex. 
One of the authors (P.~M.) would like to acknowledge support from 
 the Swiss State Secretariat for Education, Research and Innovation (SERI) - Federal Commission for Scholarships (FCS) award \#2015.0594. %
We gratefully acknowledge financial support from the Polish National Science Center,
 grants UMO-2015/18/M/ST2/00056, UMO-2016/23/D/ST2/00715 and UMO-2018/30/M/ST2/00319.  %
We acknowledge the excellent support provided by the PSI
technical groups and by various services of the collaborating
universities and research laboratories. 

\TODOdone{Please send me all further funding sources that shall appear here.}

\bibliography{blinding-refs}

\end{document}